# A Unique 10 Segment Display for Bengali Numerals

**Md. Abul Kalam Azad, Rezwana Sharmeen, Shabbir Ahmad and S. M. Kamruzzaman[†]**
Department of Computer Science & Engineering,
International Islamic University Chittagong, Chittagong, Bangladesh.
[†] Department of Computer Science & Engineering, Manarat International University, Dhaka, Bangladesh.
{azadarif, r_sharmin_79, bappi_51, smk_iiuc}@yahoo.com

**Abstract**

*Segmented display is widely used for efficient display of alphanumeric characters. English numerals are displayed by 7 segment and 16 segment display. The segment size is uniform in this two display architecture. Display architecture using 8, 10, 11, 18 segments have been proposed for Bengali numerals 0...9 yet no display architecture is designed using segments of uniform size and uniform power consumption. In this paper we have proposed a uniform 10 segment architecture for Bengali numerals. This segment architecture uses segments of uniform size and no bent segment is used.*

**Keywords**: 10 segment display, Bengali numerals, segmented display, combination vector.

## I. INTRODUCTION

Different display architectures using finite number of segments have been proposed for displaying Bengali numerals. 3X8 dot matrix system was used earlier for displaying Bengali numerals but it requires large numbers of dots to be manipulated thereby increasing the memory storage, cost and design complexity. The 7 segment display is very common in displaying English numerals and a 16 segment display is widely used for displaying English Alphanumeric characters. Modified Twin 7segment [2] is also proposed for accurate display of Bengali numerals. 8-segment [6]and 11 segment [5]display for displaying Bengali Numerals have been proposed earlier by using specially fabricated bend segments, which is quite costly and does not provide complete realistic view. 10 [4]segment and 11 [2]segments are also used, these display architectures do not use any bent segment but the segments are not of uniform size In our proposed 10 segment display we don't need to use any specially fabricated bent segments and still maintaining the quality of the accuracy and the uniformity of segment size.

## II. CHARACTERISTICS OF BENGALI NUMERALS

Bengali numerals have more curved corners in comparison to English numerals. The accuracy of display of each numeral is highly dependent on the perfect appearance of the curved corners. But the number of segments needed to represent each numeral increase linearly with the required degree of the accuracy of the curved edges [4].

## III. THE PROPOSED 10 SEGMENT DISPLAY

Our proposed uniform 10 segment display model is shown in Figure 1. Each segment is of uniform size. The segments are non-overlapping and labeled with specific alphabet.

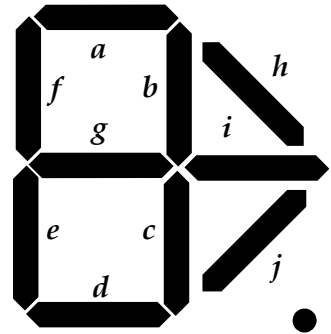

Figure 1: Proposed 10 Segment Display

## IV. REPRESENTATION OF BENGALI NUMERALS WITH THE PROPOSED 10 SEGMENT DISPLAY

The following figure, Figure 2 shows the representation different Bengali Numerals using proposed 10 segment display.

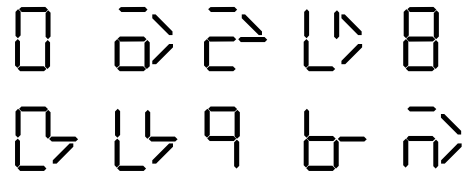

Figure 2: Proposed Segment Display of the digits

Table 1 contains the representation Bengali Numerals using 10 segment display.

Table 1: Representation of Bengali Numerals

| Digit | Bengali | Pattern | Digit | Bengali | Pattern |
|---|---|---|---|---|---|
| 0 | ০ |  | 1 | ১ |  |
| 2 | ২ |  | 3 | ৩ |  |
| 4 | ৪ |  | 5 | ৫ |  |
| 6 | ৬ |  | 7 | ৭ |  |
| 8 | ৮ |  | 9 | ৯ |  |

## V. FUNCTIONAL REPRESENTATION OF BENGALI NUMERALS

As the elements of information are 10 Bengali numerals, 4 bits are used to represent each numeral. The truth table is depicted in Table 2. After analyzing the segment activation from the truth table the logic functions and circuits are derived in section 5 and section 5 respectively.

Table 2: Truth table for the Bengali Numerals

| Digits | BCD Input | | | | Segments | | | | | | | | | |
|---|---|---|---|---|---|---|---|---|---|---|---|---|---|---|
| | w | x | y | z | a | b | c | d | e | f | g | h | i | j |
| 0 | 0 | 0 | 0 | 0 | 1 | 1 | 1 | 1 | 1 | 1 | 0 | 0 | 0 | 0 |
| 1 | 0 | 0 | 0 | 1 | 1 | 0 | 1 | 1 | 1 | 0 | 1 | 1 | 0 | 1 |
| 2 | 0 | 0 | 1 | 0 | 1 | 0 | 0 | 1 | 1 | 0 | 1 | 1 | 1 | 0 |
| 3 | 0 | 0 | 1 | 1 | 0 | 1 | 0 | 1 | 1 | 1 | 0 | 1 | 0 | 1 |
| 4 | 0 | 1 | 0 | 0 | 1 | 1 | 1 | 1 | 1 | 1 | 1 | 0 | 0 | 0 |
| 5 | 0 | 1 | 0 | 1 | 1 | 1 | 0 | 1 | 1 | 1 | 0 | 0 | 1 | 1 |
| 6 | 0 | 1 | 1 | 0 | 0 | 1 | 0 | 1 | 1 | 1 | 0 | 0 | 1 | 1 |
| 7 | 0 | 1 | 1 | 1 | 1 | 1 | 1 | 0 | 0 | 1 | 1 | 0 | 0 | 0 |
| 8 | 1 | 0 | 0 | 0 | 0 | 0 | 1 | 1 | 1 | 1 | 1 | 0 | 1 | 0 |
| 9 | 1 | 0 | 0 | 1 | 1 | 0 | 1 | 0 | 1 | 0 | 1 | 1 | 0 | 1 |

## VI. COMBINATION VECTOR FOR EACH NUMERAL

The combination vector for each numeral is shown in Table 3.

Table 3: Combination Vector for Bengali numeral

| Digit value | Bengali Numeral | Combination Vector for Bengali Numeral |
|---|---|---|
| 0 | ০ | { a, b, c, d, e, f } |
| 1 | ১ | { a, c, d, e, g, h, j } |
| 2 | ২ | { a, d, e, g , h, i } |
| 3 | ৩ | { b, d, e, f, h, j } |
| 4 | ৪ | { a, b, c, d, e, f, g } |
| 5 | ৫ | { a, b, d, e, f, i, j } |
| 6 | ৬ | { b, d, e, f, i, j } |
| 7 | ৭ | { a, b, c, f, g } |
| 8 | ৮ | { c, d, e, f, g } |
| 9 | ৯ | { a, c, e, g, h, j } |

## VII. SUM OF PRODUCT FUNCTION FOR EACH SEGMENT

The sum of product function for each is segment is shown in Table 4.

Table 4: Sum of product function for each Segment

| Segment | Sum of Product Function |
|---|---|
| a | $\sum ( 0, 1, 2, 4, 5, 7, 9 )$ |
| b | $\sum ( 0, 3, 4, 5, 6, 7 )$ |
| c | $\sum ( 0, 1, 4, 7, 8, 9 )$ |
| d | $\sum ( 0, 1, 2, 3, 4, 5, 6, 8)$ |
| e | $\sum ( 0, 1, 2, 3, 4, 5, 6, 8, 9 )$ |
| Segment | Sum of Product Function |
| f | $\sum ( 0, 3, 4, 5, 6, 7, 8)$ |
| g | $\sum ( 1, 2, 4, 7, 8, 9 )$ |
| h | $\sum (1, 2, 3, 9 )$ |
| i | $\sum (2, 5, 6, 8 )$ |
| j | $\sum (1, 3, 5, 6, 9)$ |

## VIII. MINIMIZED EXPRESSION FOR EACH SEGMENT

The minimized expression for each segment is obtained by Karnaugh-map method. The expressions are shown in Table 5.

Table 5: Minimized expression for each segment

| Segment | Minimized Expression |
|---|---|
| a | $w'y' + wz + xyz + x'yz'$ |
| b | $x + yz + w'y'z'$ |
| c | $w + y' + xyz$ |
| d | $w'x' + z' + xy'$ |
| e | $z' + x' + y'$ |
| f | $x + y'z' + yz$ |
| g | $w + xy'z' + x'y'z + x'yz'$ |
| h | $x'y + x'z$ |
| i | $yz' + xy'z + wy'z'$ |
| j | $x'z + y'z + xyz'$ |

## IX. CIRCUIT DIAGRAM

Figure 3 depicts the circuit diagram for the segments of 10 segment scheme. We have used standard inverter, 2-input, 3-input, AND and OR logic gates.

## X. USING LED AND POWER CONSUMPTION

An LED is a diode, so current will flow in one direction through it, but not in the other direction. When an LED is 'forward biased' it will light and there will be a voltage drop of around 0.7V across it. When an LED is 'reverse biased', current will not flow through it and it will not light.

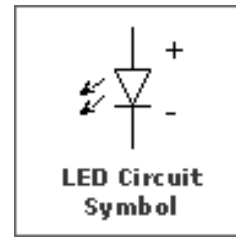

Figure 4: Led Circuit

Most LEDs require a forward bias voltage of around 2V and consume a current of around 20mA. To supply an LED with 2V at 20mA, it simply needs to place a current-limiting resistor in series with it. The resistor value can be calculated using simple Ohm's law...

$$R = \frac{V_s - 2V}{0.02A}$$

…where R is the resistor value in Ohms (Ω), and Vs is the voltage of the power supply in Volts (V). If there are different values for the forward bias voltage and current, then it should be substituted for the 2V and 0.02A values.

## XI. UNIQUENESS OF THE PROPOSED DISPLAY

LEDs rarely fail, unless they are supplied with the wrong voltage. So are handy in parts of a model that are hard to reach. In our proposed 10 segment display all the segments are of equal size hence the power consumption will also be uniform for each segment. The LED brightness for a given current can be increased by pulsing the LED at higher current. In other displays as the segments are of different sizes and some segment display also uses curved and bent segment hence with the difference in size and architecture of the segments the power consumption will also differ and the brightness of different segments will be different. So, to reduce the problems display circuits with different size of LEDs are needed to supply different amount of current which can be done by changing the value of the resistors in series with the LEDs. This made the circuits of the previously proposed segmented display complex and costly. But for our display it will be uniform for each and every segment. So there will be no different level brightness and hence there is no need to user different amount of current, which leads the circuit to be simple.

## XII. REPRESENTATION OF ENGLISH NUMERALS

English numerals can be displayed by using our proposed 10 segment display. For displaying English numerals we need not to activate the segments h, i and j. Without activating these three segments our proposed display will function like 7 segment display, hence it is possible to display the English numerals successfully.

## XIII. CONCLUSION

In our proposed 10 Segment display all the segments are of uniform size. The currently available 7 segment and 16 segment display for English alphanumeric characters have segments of equal size. The proposed segmented display is the unique segmented display that proposes a single circuit with straight segments of uniform size for displaying both Bengali and English numerals. Although we have used minimum number of segments, this display still provides appearances of digit to extended level of accuracy. So it is expected that this segment display may be used as an ideal circuit for displaying both Bengali and English numerals.